# Multi Agent Path Finding using Evolutionary Game Theory

Sheryl Paul
sherylpa@usc.edu
University of Southern California
Los Angeles California, USA

Jyotirmoy V. Deshmukh
jdeshmuk@usc.edu
University of Southern California
Los Angeles California, USA

## ABSTRACT

In this paper, we consider the problem of path finding for a set of homogeneous and autonomous agents navigating a previously unknown stochastic environment. In our problem setting, each agent attempts to maximize a given utility function while respecting safety properties. Our solution is based on ideas from evolutionary game theory, namely replicating policies that perform well and diminishing ones that do not. We do a comprehensive comparison with related multiagent planning methods, and show that our technique beats state of the art RL algorithms in minimizing path length by nearly 30% in large spaces. We show that our algorithm is computationally faster than deep RL methods by at least an order of magnitude. We also show that it scales better with an increase in the number of agents as compared to other methods, path planning methods in particular. Lastly, we empirically prove that the policies that we learn are evolutionarily stable and thus impervious to invasion by any other policy.

## 1 INTRODUCTION

Large-scale deployment of multiagent robotic systems is fast becoming a reality with applications in warehouse settings, firefighting, search and rescue missions in disaster recovery scenarios, and exploration of unknown environments. *Multiagent path finding* (MAPF) [38] is an important challenge problem for such systems that involves finding and planning suitable paths that ensure that each agent accomplishes its goal while avoiding collisions. MAPF has been applied to various problems, such as surveillance and patrolling [9, 16, 26], warehouse settings [2, 40], coordinating tug robots [23], train and aircraft scheduling [21, 27], among others. The initial work on MAPF assumes that the environment is fully known, the planner is centralized and can fully observe the environment and the locations of the agents therein, and that each agent explicitly knows the goal locations. This allows early MAPF approaches to use informed graph search algorithms to efficiently obtain the optimal collision-free paths for agents. Later approaches investigate decentralized and hierarchical planning approaches. [6]

In this paper, we wish to address the multiagent path finding problem in environments that can have stochastic uncertainties, where agents are fully autonomous, and where the agents may not know the goal locations *a priori*. There are several practical scenarios where our algorithms could be applicable: for example, consider a warehouse setting where robots move items from shelves to the closest available empty container, or wildfire monitoring drones that need to find a path to the closest dangerous patch of wildfire. Approaches such as heuristic A* search and its variants [12] work very well in such scenarios for single agent systems; however, as we demonstrate in this paper, the scalability of naïve A* search-based methods suffers with increasing number of agents and the size of the environment.

Our solution is based on techniques from evolutionary game theory (EGT) [36]. Evolutionary game theory attempts to model the behavior of large populations who repeatedly engage in strategic interactions [29]. Populations follow the Darwinian principle of survival and growth based on some fitness function, and we can dynamically model the success of chosen strategies over time. In our setting, we assume that the environment is stochastic and that the agents are autonomous and homogeneous. Here, by homogeneity, we mean that all agents utilize the same learned policy. We also assume that agents do not know the exact locations of their goals, and that they can start in any permissible start location in the environment. We assume that the state of the agent depends on its observation space (as defined by its perception sensors)[1]. We assume that the user provides a utility function that can be evaluated on a specific path chosen by the agent. The goal of our method is to obtain a stochastic policy that maximizes the expected value of the utility function across all agents irrespective of the start location of each agent. The key idea is to treat the space of all possible state-action pairs as a population, where state-action pairs that are involved in high utility paths have increased probability of replication.

With recent advances in planning for uncertain environments, many related methods [1, 44] that solve similar problems have emerged. Model-free Reinforcement learning (RL)-based methods [24, 41] use episodic experiences learned by a set of agents interacting with the environment to obtain policies that maximize a suitably defined reward function [15]. There is a vast amount of literature dealing with the theoretical aspects of multiagent RL (see [4, 47] for details.)

Multiagent systems can also be viewed from a game-theoretic lens, and there has been recent interest in pursuing game-theoretic methods for path planning [46]. Such methods typically focus on agent interactions and obtain optimal policies by solving extensive form games [10] or focus on reward-shaping using mechanism design [19, 35]. Recently, there have been some cross-overs between game-theoretic methods and RL, in approaches that use solution concepts like the Shapley value for cooperative games to serve as rewards in RL [11, 14].

**Contributions:** Our main contributions are as follows:

(1) We investigate the problem of multiagent path finding (MAPF) in stochastic and previously unknown environments, where agents are homogeneous and have limited visibility.
(2) We permit agents to have multiple possible start locations and for the environment to provide multiple possible goal locations.

---

[1]In our experimental results we assume that agents have no visibility beyond the grid location that they occupy, but our framework permits agents that can have visibility extending to multiple cells or can also depend on communication with other agents.

(3) We develop an algorithm based on evolutionary game theory to solve MAPF. We show that our technique is able to find shorter paths than RL algorithms such as the Monte Carlo method [39], Q-learning [42] and a deep RL method based on proximal policy optimization (PPO) [31].
(4) We show that our algorithm leads to faster training times compared to Q-learning and PPO, and scales better with increasing number of agents compared to path planning methods like A*.
(5) Our algorithm requires fewer policy updates than the Monte Carlo method or Q-learning, and is more space-efficient than methods that use batchwise gradient descent (e.g., some implementations of PPO).

## 2 PRELIMINARIES

In this section, we present the basic terminology and notation required to define the problem that we wish to solve. We consider a multi-agent system consisting of $N$ autonomous agents that interact with each other within an environment. We model the environment as a finite, *rewardless*, Markov Decision Process (MDP), formally defined as follows:

*Definition 2.1 (Rewardless MDP).* A rewardless MDP is a tuple $(S, A, \Delta, \iota)$ where $S$ is a finite set of states, $A$ is a finite set of actions, $\Delta$ is a set of transitions of the form $(s, a, p, s')$, where $s, s' \in S, a \in A$, and $p \in [0, 1]$ indicates the probability of executing that action, and $\iota : S \to [0, 1]^S$ is a function that maps each agent state to the probability of being an initial state ($\sum_s \iota(s) = 1$).

Let $\mathcal{A}$ denote the set of agents, $\mathcal{A}_i$ denote the $i^{th}$ agent, and $t$ denote an arbitrary time of interest at which we observe the behavior of the environment. We assume that we are interested in *episodic behaviors* of an agent interacting with the environment, i.e., we assume that agent behaviors start at time 0 and end at a finite time $T$. In other words the length of the episode is bounded above by $T$. The state of agent $\mathcal{A}_i$ at time $t$ is denoted as $s^i_t$. The temporal behavior of the agent is consistent with the structure of the environment MDP, i.e., if the agent is in state $s^i_t \in S$ at time $t$ and it takes an action $a^i_t$, then it stochastically moves to some state $s^i_{t+1}$ with probability $p$.

*Definition 2.2 (Stochastic Policy).* A stochastic policy $\pi$ for an MDP is a probability distribution over its actions conditioned by the current state, denoted as $\pi(a \mid s)$.

Once we fix a stochastic policy, the MDP essentially becomes a Markov chain, where at any state $s^i_t$, the $i^{th}$ agent executes the action $a^i$ sampled from from $\pi(a \mid s = \mathcal{A}^i_t)$. In general, each agent may have its individual policy, but, in this paper, we assume that agents are homogeneous, i.e., they all share the same policy.

### 2.1 Problem Definition: Goal-seeking policies

An *MDP with goal states* is simply a rewardless MDP augmented with a set of *goal* states $\Phi \subset S$. In order to formally define our problem, we first introduce the notion of state-action paths (or simply paths) under a fixed policy.

*Definition 2.3 (State-Action Paths, Path Length, Final State).* Given a policy $\pi$, a trajectory or path for the $i^{th}$ agent under $\pi$ (denoted $\tau^i(\pi)$) is defined as an alternating sequence of states and actions defined as follows:

$$\tau^i(\pi) = \left(s^i_0, a^i_0, \ldots, s^i_{\ell-1}, a^i_{\ell-1}, s^i_\ell\right) \quad (1)$$

Here, $s^i_0$ is sampled from $S$ with probability $\iota(s^i_0)$. The time index of the last state in the sequence is at most the time horizon $T$, i.e. $\ell \leq T$. For each $j \in [0, \ell - 1]$, the action $a^i_j$ is sampled from the stochastic policy $\pi(a \mid s)$ conditioned on $s = s^i_j$, and for all $j \in [0, \ell - 1]$, there is a $p > 0$ s.t. the tuple $(s^i_j, a^i_j, p, s^i_{j+1}) \in \Delta$.

We denote that the path is sampled from the policy using the notation $\tau \sim \pi$. For a path as shown in (1), the *path length* (denoted $|\tau|$) is equal to $\ell + 1$ (number of states in the path). The last state in the trajectory is denoted as $final(\tau^i(\pi))$; in (1), this is $s^i_\ell$.

We are now ready to define the problem that we wish to solve in this paper. Informally, we wish to find policies such that all agents starting from a randomly chosen initial state can reach some goal state within the maximum permitted time horizon $T$. Formally,

$$\pi^* = \arg\max_\pi \min_{i \in \{1, \ldots, N\}} \Pr\left(final(\tau^i(\pi)) \in \Phi \land |\tau^i(\pi)| \leq T\right) \quad (2)$$

In other words, we want to find a policy that maximizes the probability that every agent reaches a goal state in less than $T$ number of steps.

## 3 SOLUTION APPROACH

In this section, we describe an algorithm based on the principles of *evolutionary game theory* (EGT) to also solve the problem stated in (2). We first do a brief review of EGT and related notions from game theory.

The early work on EGT has roots in the application of traditional game theory to biological populations [36]. EGT utilizes Darwinian principles of the fittest surviving and replicating, thereby growing in population while the weaker populations diminish. It differs from classical game theory in that it focuses more on the change in population dynamics over the passage of time, where a specific population group would be classified as ones following the same strategy. We now describe how we apply EGT; the first step is to encode the problem of learning goal-seeking policies using dynamical evolution of populations.

**Encoding policies as populations**

We consider our population to consist of $S \times A$ organisms. We sample a policy $\pi$ consisting of a set of organisms for replication. In order to choose which organisms to replicate, we define a fitness function over the trajectory induced by the policy. We can define this fitness function as:

$$u = \frac{L}{d(s_0, s_L)}$$

where $d(s_0, s_L)$ is the Manhattan distance between the locations in states $s_0$ and $s_L$. We assign constructpolicy as a piece-wise function that is implementation dependant, for the population to be replicated i.e. increased if the fitness function is high, or decreased if it is low.

After a sufficient number of episodes, we can construct a stochastic policy $\pi_{trained}$ using the algorithm ConstructPolicy.

An 'Evolutionarily Stable Strategy' (ESS) is defined as a strategy (or policy), that cannot be invaded by the introduction of a foreign strategy ($\pi_{new}$) with a small enough probability $p_{new}$ where $0 \leq p_{fnew} \leq 1$.
$$u(\pi_{trained}) \geq p_{new}(u(\pi_{new})) + (1 - p_{new})(u(\pi_{trained}))$$

**Algorithm** MAPF-EGT algorithm

**Initialize:** $\pi_{rand}(a|s) \leftarrow \frac{1}{|A|}$, $counter(s,a) \leftarrow \bot, \forall s \in S, a \in A, j = 0$
**for** e = 1 to Number of Episodes: **do**
   **Initialize:** $\tau^i = ()$ ▷ Initialise empty trajectory
   Sample: $s^1{}_0, \ldots, s^i_0 \sim \iota$ ▷ Sample initial states
   **while** $j \leq T$ **do**
     **for** all $i \in \{1, \ldots, N\}$ **do**
       **if** $s^i{}_j \notin \Phi$ **then** ▷ While agent has not reached the goal
         $a^i_j \sim \pi_{rand}$ ▷ Sample action from random policy
         $\tau^i = \tau^i \circ (s^i{}_j, a^i{}_j)$ ▷ Append (s,a) pair to trajectory
         $(s^i{}_j, a^i{}_j, s^i{}_{j+1}) \sim \Delta$ ▷ Sample next state from transitions
       **end if**
       **if** $s^i{}_{j+1} \in \Phi$ **then**
         update($\tau^i_{j+1}$) ▷ Perform update if agent has reached
       **end if**
     **end for**
     $j \leftarrow j + 1$
   **end while**
   **for** all $i$, where $s^i{}_T \, not \in \Phi$ **do**
     update($\tau^i_{j+1}$) ▷ Perform update for all agents that have not reached
   **end for**
**end for**

We describe our algorithm as follows: Initially our policy is defined to be random, and our counter(s, a) is defined to be $\bot$ for all state-action pairs. For each episode, we initialise an empty trajectory for all agents. While the episode does not terminate, we sample an action from the policy and append the state-action pair to the trajectory for each agent that has not yet reached the goal. We then move to the next state according to the structure of the MDP. If the agent reaches the goal we update the trajectory. At the end of the episode, the trajectories of all agents that have not reached the goal are also updated. The update criteria are implementation dependent and will be described in section 4.2.

## 4 EXPERIMENTAL SETUP

### 4.1 Baselines for comparision

Traditionally, the definition of an MDP also includes state-based or transition-based *rewards*. Formally, a transition-based reward function $R$ maps a state in $s$ and an action in $A$ to some number in $\mathbb{R}$. Reward functions allow us to define the *return* or *payoff* for a sequence of actions performed by the agent.

In contrast to reward function-based RL methods, we define the notion of path fitness.

*Definition 4.1 (Return or Payoff using Reward Functions).* The return for any agent $i$ at time $T$ can be defined as the sum of its rewards over an episode
$$G^i_T = \sum_{t=0}^{T} R(s^i_t, a^i_t)$$
where $a^i_t$ is sampled from $\pi(a|s^i_t)$.

**Algorithm** ConstructPolicy

1: **for** $s \in S$ **do**
2:   **if** $\forall a$, $counter(s,a) == \bot$ **then**
3:     $\pi_{train}(a|s) = \frac{1}{|A|}$
4:   **else**
5:     $sum = \Sigma_{a \in A} \, counter(s,a) \neq \bot$
6:     **for** $a \in A$: **do**
7:       **if** $counter(s,a) \leq 0$ **then**
8:         $\pi_{train}(a|s) = \bot$
9:       **else**
10:         $\pi_{train}(a|s) = \frac{counter(s,a)}{sum}$
11:       **end if**
12:     **end for**
13:   **end if**
14: **end for**
15: $\pi_{final}(a|s) = (1-\epsilon)(\pi_{train}(a|s)) + (\epsilon)(\pi_{rand}(a|s)), \forall s \in S, a \in A$

The return in a multi-agent setting can be defined as an appropriate combination of the returns for all agents at time $T$.

For example, the cumulative return defined below is the sum of all returns, but overall returns based on maximum return or average return per agent could also be defined.
$$G_T = \oplus_i \sum_{t=0}^{T} R(s^i_t, a^i_t).$$

In order to follow the collision-avoidance nature of path-finding we introduce sets of permissible actions in each state. We define $E : S \rightarrow 2^A$.

If $(s, a, p, s') \in \Delta$, and $p > 0$:
$$s \neq s' \implies a \in E(s)$$
$$\text{and } a \notin E(s) \implies s = s'.$$

In other words, agents taking impermissible actions in a state return do not change their state. We now define our rewards as follows:

$$R(s, a, s') = \begin{cases} \delta_1 \text{ if } s' \notin \Phi \\ \delta_2 \text{ if } a \notin E(s) \\ \delta_3 \text{ if } s' \in \Phi \end{cases} \quad (3)$$

where $\delta_2 < \delta_1 < 0 < \delta_3$

LEMMA 4.2. *Policies that maximize rewards as defined above solve the problem defined in* (2).

PROOF. Proof follows from the fact that policies that maximize expected payoff must receive a positive reward, and the only way to receive a positive reward is by reaching a goal state. □

**MAPF using A\* search:** A\* is an informed search problem which means the planner knows the goal states *a priori*. It maintains a tree of paths that originate at the start state and extends these paths by expanding one state and generating all of its successors at each iteration until it reaches the goal. It chooses which state to expand by finding the state which minimizes :
$$f(s) = g(s) + h(s)$$

where $s$ is the next state on the path, $g(s)$ is the cost of travelling to state $s$ from the start state and $h(s)$ is the heuristic of the cost it will take to travel from state s to the goal state. It has been shown that A* finds the shortest path with a heuristic that does not overestimate the distance to the goal state.

There are many variants and extensions of A* such as Incremental A*[17], D*[37], D* lite[18], Iterative Deepening A*[20], etc. We will simply focus on A* as our baseline.

***Monte Carlo Reinforcement Learning:*** Unlike A*, Monte Carlo[34] based methods do not require any prior knowledge of the environment, and solely rely on previous episodes or simulations to estimate state/state-action based rewards.

$$v_\pi(s) = \mathbb{E}_{s'|(s,a,s')\sim\Delta}(R(s,a,s') + v_\pi(s'))$$

In particular, in order to estimate $v_\pi(s)$ (i.e. the value of a state s) given a set of episodes E generated by following the policy $\pi$ and passing through s: We call each occurrence of s in an episode a *visit* to s, $G_j$ is the return in episode j after visiting s, and G is the set of all such rewards. Then:

$$v(s) = Average(G)$$

The first visit Monte Carlo approach estimates $v_\pi(s)$ as the average of returns following the first occurrence of state s. We will use this approach to compare against our own.

***Q learning:*** Q-learning [39] is a model-free RL algorithm that maximizes the expected value of the total reward over any and all consecutive states following the starting state. The algorithm maintains a table of approximate Q-values for each state-action pair. The Q table comprising of $Q(s,a) \forall s \in S, \forall a \in A$ is initialised to random values at the beginning. After each step, it is updated as:

$$Q^{new}(s,a) \leftarrow Q^{current}(s,a) + \alpha(R(s,a,s')$$
$$+ \gamma \max_a Q(s',a) - Q^{current}(s,a))$$

Here, $\alpha$ is the learning rate ($0 \leq \alpha \leq 1$) which determines the weight of the new information over the old, and $\gamma$ is the discount factor ($0 \leq \gamma \leq 1$) which values the rewards received later, lower than the rewards received presently.

***Proximal Policy Optimization:*** The exact details of the PPO algorithm can be found in [31]. PPO is a policy gradient algorithm that assumes that a policy that is parameterized; for example, the policy could be represented by a neural network, and the set of weights of the neural network comprise the policy parameters. The key idea in PPO is to ensure that subsequent updates to the policy (through a stochastic gradient of the objective function) do not cause drastic changes in the policy. The objective function used by PPO is proportional to the episodic return as well as the ratio of the probability under the new and old policies.

## 4.2 Description of Environment used for Experiments

For our experiments, we define the world as a two dimensional $n \times n$ grid, where each state $s \in S$, is defined as $(x,y)$ where $x$ and $y$ are coordinates. We define $O, G \subset S$, O is the set of obstacle states such that $s_t^i \nsubseteq O, \forall i \in N, 0 \leq t \leq T$ and G is the set of goal states. The algorithms aim to find a path from $s_0^i$ to any state $s \in G$ that maximises the total reward across all agents in an episode. The set of actions $A = \{up, down, left, right, stay\}$.

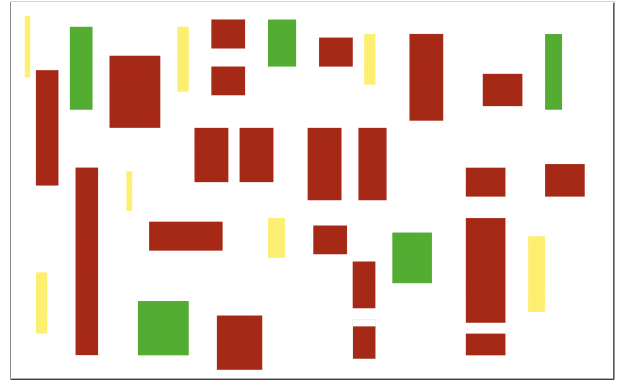

**Figure 1: Sample Grid World**
**Yellow: Start Locations, Red: Obstacle Locations, Green: Goal Locations**

Further we specify the update function in our algorithm for our experimental setup.

**Algorithm** UpdateFunction
---
1: update($\tau_\ell^i$)
2: $u = \frac{l}{d(s_0, s_\ell)}$
3: **if** $s_\ell \in G$ **then**
4:   **if** $u \leq \eta$ **then**      ▷ where $0 \leq \eta \leq 1$
5:     $update\_prob = 1 - (1-u)^\alpha$    ▷ update for shorter paths
6:      ▷ where $\alpha > 1$
7:   **else**
8:     $update\_prob = 1/u$    ▷ update for longer paths
9:   **end if**
10:   **if** $rand(0,1) \geq update\_prob$ **then**
11:     $counter(s_k, a_k) + \nu$ ▷ positive update for (s,a) pairs in shorter paths    ▷ where $\nu \in \mathcal{I}^+$
12:   **end if**
13: **else**
14:   **if** $u \geq \beta$ **then**
15:     $counter(s_k, a_k) - \mu$ ▷ negative update for (s,a) pairs in longer paths    ▷ where $\mu \in \mathcal{I}^+$
16:   **end if**
17: **end if**

Our research questions are as follows:

**Research Hypothesis 1:** EGT for MAPF does as well or better than A*, Monte Carlo RL, Q Learning and PPO in minimising the average path length from the initial to the goal location with increasing size of the world and number of agents.

**Research Hypothesis 2:** EGT for MAPF requires comparable to or less computation time than A*, Monte Carlo RL, Q Learning and PPO to find the policy $\pi*$ that maximizes the average reward with increasing size of the world and number of agents.

**Research Hypothesis 3:** EGT for MAPF requires fewer policy updates than Monte Carlo RL and Q Learning to find the policy $\pi*$ that maximizes the average reward with increasing size of the world and number of agents.

## 5 EXPERIMENTAL EVALUATION AND RESULTS

We collect results for increasing grid sizes, and number of agents for each of the aforementioned algorithms.

Our experiments were carried out on a laptop with 2.0GHz dual-core Intel Core i5 processor and 16GB RAM.

Q Learning and PPO perform updates step-wise and thus require much longer to train. As opposed to MAPF-EGT, which performs updates on only a selected number of episodes, these algorithms also acquire more training data in fewer episodes. We time out these algorithms accordingly.

We use the 'stablebaselines3' [28] implementation of PPO with standard values for the hyperparameters. Our algorithms have been implemented on OpenAi gym [3].

### 5.1 Experiments with increasing grid size

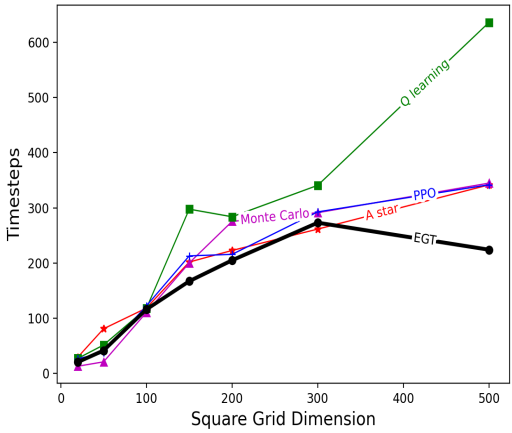

Figure 2: Comparison of average best path length with other algorithms with increasing grid size

*Comparison of path length (timesteps):* The experimental results indicate that MAPF-EGT performs comparatively better than the other algorithms in terms of path length (in timesteps) for large grids. PPO and Monte Carlo are comparable to or even slightly better than MAPF-EGT for smaller grid sizes but perform worse as the dimensions increase. Q learning does not scale well as it requires much longer to train in order to perform comparatively well. Due to inadequate training, Q learning has a lower success rate as compared to the other algorithms for large grids.

*Comparison of computation time:* We also compare computation time, which includes the training time and run time. MAPF-EGT performs better in terms of computation time in comparison to the other algorithms for large grids except for A*. However we see later that A* scales badly for increase in the number of agents. Monte Carlo RL is faster for smaller grids (upto 200x200), but slows down thereafter. We limit the training time based on the size of the grid and the time it requires to train other algorithms for the same grid. The computation time for Q learning grows exponentially and remains insufficient for an adequate number of training

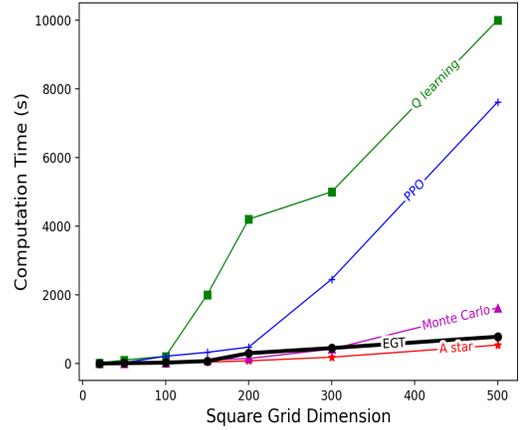

Figure 3: Comparison of computation time against existing algorithms with increasing grid size

episodes to achieve good performance. The highest time required by MAPF-EGT was only 13 minutes.

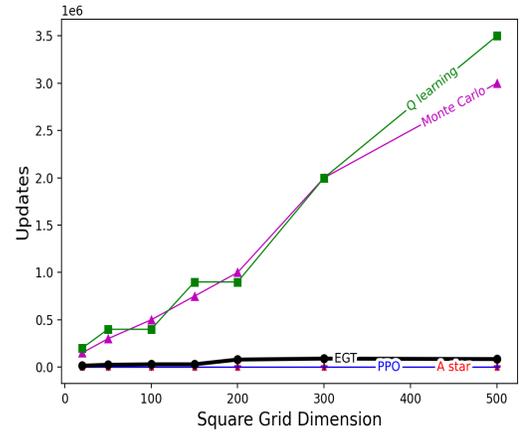

Figure 4: Comparison of Policy Updates against existing algorithms with increasing grid size

*Comparison of number of policy updates:* Another comparison we make is the number of updates to the policy during the training. MAPF-EGT and the Monte Carlo method both perform updates episodically, while Q learning and PPO perform updates step-wise or in batches of steps. However, unlike the other methods, MAPF-EGT considers only some episodes with performance sufficiently good or bad for updating its policy. This drastically reduces the number of policy updates to between 15,000 and 90,000. The Monte Carlo method on the other hand, with the same number of training episodes performs anywhere between 75,000 to 3 million updates which is nearly two orders of magnitude more than MAPF-EGT. Q learning ranged from 200,000 to a worst case scenario of 35 million updates, nearly three orders of magnitude higher than MAPF-EGT. As PPO performs only batches of updates, where every batch consisted of 2048 timesteps, it performs the lowest number of updates upto 250 in the worst case. If extended to continuous spaces, the size of the neural network to represent the environment would grow extremely large, which might cause batch-wise updates to slow down, and require more memory. A star does not perform any updates and we indicate the number to be zero in the graph.

| Algorithm | Metric for Comparison | Grid Size | | | | | | |
|---|---|---|---|---|---|---|---|---|
| | | 20x20 | 50x50 | 100x100 | 150x150 | 200x200 | 300x300 | 500x500 |
| A star | Timesteps | 29.7 | 81.0 | 118.41 | 201.85 | 222.77 | 261.48 | 341.67 |
| | Expected min distance from obstacles | 2.34 | 2.36 | 12.17 | 10.19 | 13.42 | 13.15 | 10.81 |
| | Computation Time (s) | 1 | 5 | 9 | 41 | 77 | 184 | 537 |
| | Policy Updates | 0 | 0 | 0 | 0 | 0 | 0 | 0 |
| Monte Carlo | Timesteps | 13.19 | 21.11 | 110.59 | 200.19 | 275.37 | 291.23 | 344.79 |
| | Expected min distance from obstacles | 1.19 | 1.08 | 2.06 | 2.96 | 1.23 | 1.48 | 1.07 |
| | Computation Time (s) | 2 | 4 | 20 | 100 | 145 | 420 | 1620 |
| | Policy Updates | 150000 | 300000 | 500000 | 750000 | 1000000 | 2000000 | 3000000 |
| Q learning | Timesteps | 27.39 | 51.61 | 117.85 | 297.57 | 283.79 | 340.73 | 635.88 |
| | Expected min distance from obstacles | 1.5 | 0.34 | 7.07 | 11.22 | 14.49 | 13.82 | 11.35 |
| | Computation Time (s) | 10 | 100 | 210 | 2000 | 4200 | 5000 | 10000 |
| | Policy Updates | 200000 | 400000 | 400000 | 900000 | 900000 | 2000000 | 3500000 |
| PPO | Timesteps | 26.79 | 37.45 | 122.43 | 212.98 | 215.72 | 292.64 | 341.75 |
| | Expected min distance from obstacles | 1.86 | 1.27 | 5.0 | 8.01 | 11.65 | 8.61 | 7.75 |
| | Computation Time (s) | 13 | 24 | 212 | 325 | 477 | 2455 | 7614 |
| | Policy Updates | 4 | 4 | 13 | 50 | 50 | 150 | 250 |
| EGT | Timesteps | 20.75 | 41.62 | 116.05 | 167.13 | 205 | 273.06 | 224.12 |
| | Expected min distance from obstacles | 1.22 | 1.36 | 5.01 | 5.97 | 6.94 | 6.73 | 6.94 |
| | Computation Time (s) | 2 | 8 | 25 | 72 | 300 | 450 | 780 |
| | Policy Updates | 15000 | 25000 | 30000 | 30000 | 80000 | 90000 | 85000 |

**Table 1: Comparison of MAPF-EGT against existing algorithms with increasing grid size**

*Comparison of expected minimum distance from obstacles :* Expected minimum distance from obstacles is a safety measure which checks for proximity to obstacles, edges and other agents, so a higher score is desirable. For RL algorithms like Q learning and PPO where disabled actions are penalised more than permissible actions, the value for the states with illegal actions will be lower, and this will propagate to states around them with the Bellman Equation. This gives some measure of robustness with respect to the distance from cells with disabled actions. Currently a drawback of our method is that the utility function does not take this into account. As our utility function does not take into consideration the distance from obstacles, we see that MAPF-EGT generally maintains a sufficient distance and is comparable to but outperformed by some existing algorithms.

## 5.2 Experiments with increasing number of agents

In order to test the scaling with the number of agents for MAPF-EGT against other algorithms, we fix the grid size to 100x100 and vary the number of agents from 2 to 100. We also trained Q learning and PPO for more timesteps than we did in the previous experiment for comparable performance (in path length), to other algorithms. All algorithms achieved average path lengths from 90-115. We observe the Q learning take the longest time to train, but has the lowest run time of all the algorithms. It beats MAPF-EGT in total time, but we expect this will not be the case for larger grids as its training time grows exponentially. A star requires no pre-computation and has the same run time as its total time. The frontier for multiple agents grows large with increase in the number of agents and we expect this is why A star does not scale as well as the other algorithms. PPO scales comparably to MAPF-EGT but surprisingly, the Monte Carlo method does slightly worse. We note that in MAPF-EGT and Monte-Carlo, the training times are so small that the total time is nearly the same as the run time.

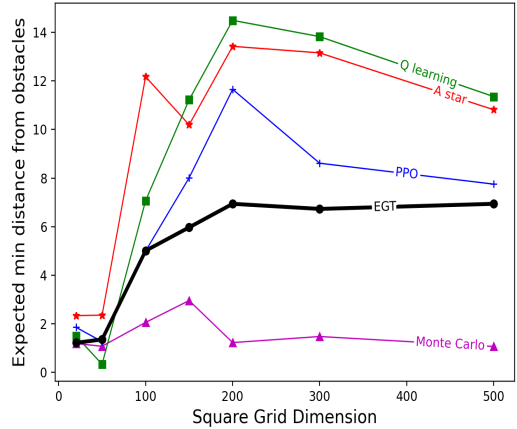

**Figure 5: Comparison of expected minimum distance from obstacles with existing algorithms with increasing grid size**

| Algorithm | Measure | Number of Agents (100x100 grid) | | | | | | | |
|---|---|---|---|---|---|---|---|---|---|
| | | 2 | 5 | 10 | 20 | 30 | 40 | 50 | 100 |
| A Star | Total time | 22 | 62 | 125 | 275 | 361 | 468 | 593 | 1719 |
| Monte Carlo | Total time | 28 | 50 | 96 | 203 | 342 | 523 | 751 | 1172 |
| | Run Time | 11 | 33 | 79 | 186 | 325 | 506 | 734 | 1155 |
| Q Learning | Total time | 593 | 595 | 597 | 602 | 608 | 613 | 621 | 655 |
| | Run Time | 1 | 3 | 5 | 10 | 16 | 21 | 29 | 63 |
| PPO | Total time | 916 | 957 | 1017 | 1057 | 1118 | 1196 | 1215 | 1756 |
| | Run Time | 16 | 57 | 117 | 157 | 218 | 296 | 315 | 856 |
| MAPF-EGT | Total time | 40 | 50 | 102 | 175 | 227 | 325 | 393 | 826 |
| | Run Time | 15 | 25 | 77 | 153 | 202 | 301 | 368 | 801 |

Table 2: Comparison of MAPF-EGT against existing algorithms with increasing number of agents

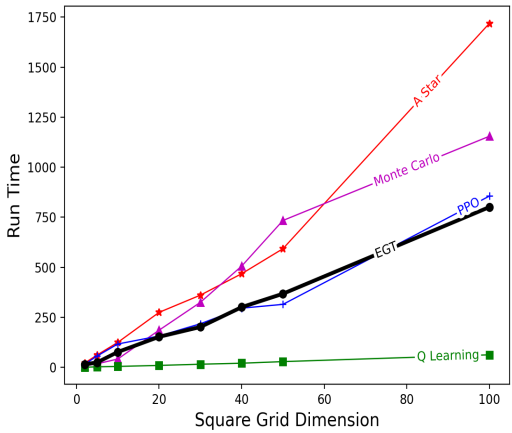

Figure 6: Comparison of run time against existing algorithms with increasing number of agents

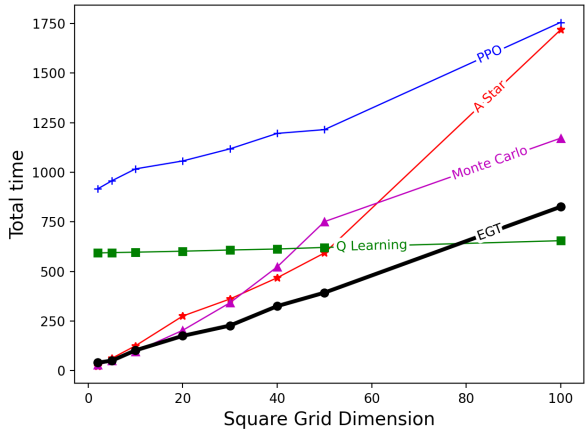

Figure 7: Comparison of total time (training time + run time) against existing algorithms with increasing number of agents

***Convergence to ESS***. In order to show that MAPF-EGT converges to an ESS, we capture the evolution of policies over time. Fig 8 shows that the heat map for all actions at time t0 is the same, t1 indicates the middle of the training process. t2 is taken around the end of the process. After introducing some number of random runs(>10% of total training runs)/updates to the policy, in t3 we see that there is still no change in the policy. This indicates that our learnt policy is resilient to the invasion by another other introduced policy and is therefore an ESS.

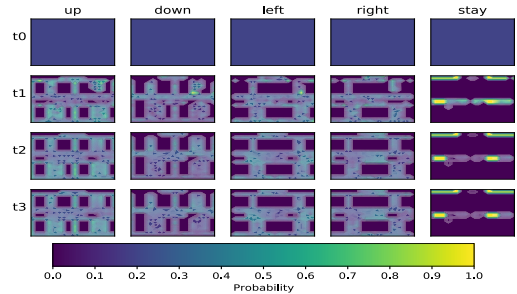

Figure 8: Evolution of policies over time

Therefore, we see from our results that all of the research hypotheses in the above section have been satisfied.

## 6 RELATED WORK

Multi Agent Path Finding (MAPF) and Multi Agent Reinforcement Learning (MARL) can be broadly classified into centralized and decentralized approaches. Centralized approaches learn a single joint policy for all the agents, suffering from the disadvantage of the curse of dimensionality [30]. Many joint approaches fail as the state-action space expands combinatorially, and this requires impractical amounts of training data. Decentralized learning involves each agent autonomously learning its own policy, which could lead to instability, even in simple matrix games[45]. An in-between is 'Centralized Training Decentralized Execution' (CTDE) which can be likened to an Actor-Critic approach. A centralized critic takes in all of the input and generates the policy, while the actors (i.e. agents) act on this information individually.

*Actor Critic Based Approaches.* The use of PPO based on the CTDE approach is studied in several works such as MAPPO [45]. They empirically show how PPO can be extended to the multi agent setting and performs comparably to offline algorithms such as Multi Agent Deep Deterministic Policy Gradient (MADDPG)[25] with minimal tuning of hyperparameters. They also identify five implementation and hyperparameter factors that strongly influence the performance of PPO and comment on how they can be optimized. While this work provides significant insights toward minimizing tuning in PPO, the need to tune at least 5 hyperparameters and to train extensively for optimal performance remains a limitation.

The approach in MAPPER [24] also involves the use of the principles of evolutionary algorithms by performing crossover, mutation and selection on the model parameters of each agent so that the probability of updating weights is directly proportional to the accumulated reward of the agent. It performs well where the grid size is limited and is relatively uncomplicated with dynamic obstacles. This algorithm is also limited in its requirement of extensive training to achieve good performance. Additionally, this algorithm penalizes oscillation by keeping a history of the states which is inconsistent with the Markov property.

The authors of [41] use a CTDE multi-agent actor-critic approach consisting of multiple actors and a single critic with heterogeneous agents. In order to combat the difficulty of working with heterogeneous agents, they introduce an embedding network similar to the work in [43]. It is a fully connected layer that is assigned to each actor which transforms observations from each actor into a fixed length vector. This allows parameter sharing similar to a setting with homogeneous agents. The authors in [7], adopt PPO with modifications for multiple agents to reach multiple goals. They optimize the convex combinatorial problem of minimizing wait time and distance travelled by empty cabs. The experiments, run on real data from New York City, show very good results at the cost of extensive training data (on the order of tens of millions of episodes), as well as large training and wait times. The ride hailing problem has also been studied from the perspective of temporal difference learning and the bandit problem in [5]. The work in [48] also uses MARL along with reward function design to solve the problem of minimising wait times at traffic signals.

*Game Theoretic Approaches.* The work by [10] also tries to empirically show the achievement of an ESS in a game of foraging and acquiring resources set in a 2D grid space. We note that the game only consists of the interaction between agents in close proximity. The limitation of the results is that this algorithm is only compared against random strategies and a limited version of itself. The performance is not adequately better than random.

The authors of [35] work on the problem of incentivizing co-operation among selfish agents where collaboration might be beneficial. Their approach involves mechanism design. Another multi agent game theoretic approach in a different setting is seen in [13]. They model multiagent interaction as correlated games where agents take decisions based on the partial information available to them. The authors of [8] work on the problem of coordinating multi robot swarms by formulating their interactions as extensive form games.

*Path Planning Approaches.* We also examine a number of works based on Conflict Based Search [33]. The algorithms proposed by [2] work on dynamic environments with pre-decided constraints. The two algorithms - parallel and abstract CBS- are extensions of Conflict Based Search. They also theoretically prove that the search methods are complete and optimal. A limitation is that the environment is learned by training on possible fixed configurations, rather than being dynamic in online settings.

The work in [22] consists of other extensions of CBS, namely Multi Constraint CBS (MC-CBS) and Large Agent CBS (LA-CBS). This work corrects the usual assumption that agents do not necessarily occupy single points on the grid, rather requiring more complexity in maneuvering. These algorithms empirically outperform classic CBS by 3 orders of magnitude. The work in [32] presents an algorithm: LC-MAE (local multi agent evacuation). The empirical results are successful and efficient for a number of 2D spaces that differ in the level of navigation. CBS approaches, like A* and other informed search algorithms, are heavily dependent on the heuristic used. The authors in [1] also use a heuristic based approach to solve a multi agent patrolling problem where agents need to visit locations with specific frequencies in an environment that is hard to navigate.

## 7 CONCLUSIONS AND FUTURE WORK

We study the problem of Multi Agent Path Finding in unknown environments with generalizable initial locations. We use an evolutionary game theoretic approach to develop our algorithm: MAPF-EGT. We compare it to existing approaches in terms of scalability in environment size and number of agents. We use metrics such as computation time, path length and number of policy updates. We note that our algorithm performs comparable to or sufficiently better than these approaches under some constraints. A limitation is that our utility function does not support safety measures for minimizing distance from obstacles. We propose using Signal Temporal Logic properties in future work to add robustness to our algorithm. Our policy update function is theoretically parallelizable, and we propose an extension to parallelize the training to speed it up. We will also work on extending to continuous and dynamic environments.